\documentclass[11pt,a4paper]{article}
    \usepackage{jheppub}

\usepackage{amsmath, amsfonts, amsthm, amssymb, graphicx}
\usepackage{color}
\usepackage[utf8]{inputenc}
\usepackage{lipsum}
\usepackage{centernot}
\usepackage{ulem}
\usepackage{feynmp}
\usepackage{pifont}
\newcommand{\cmark}{\ding{51}}%
\newcommand{\xmark}{\ding{55}}%

\usepackage{tensor}

\def\be{\begin{equation}}
\def\ee{\end{equation}}
\def\I{{\cal I}}

\newcommand{\citeseq}{\cite{KP1,KP2,KP3,KPSZ, KPS, KP4,etude,irat,mon}}
\newcommand{\citecc}{\cite{zeldovich,wein,pol,cliff,me}}
\newcommand\myeq{\mathrel{\overset{\makebox[0pt]{\mbox{\normalfont\tiny\sffamily def}}}{=}}}

\title{Cosmological consequences of {\it Omnia Sequestra}}
\author[a]{Ben Coltman}
\emailAdd{Ben.Coltman@nottingham.ac.uk}
\affiliation[a]{School of Physics and Astronomy, University of Nottingham, Nottingham NG7 2RD, United Kingdom}
\author[b,a]{Yixuan Li}
\emailAdd{yixuan.li@ens-paris-saclay.fr}
\affiliation[b]{\'Ecole normale sup\'erieure Paris-Saclay, Universit\'e Paris-Saclay, 61 Avenue du Pr\'esident Wilson, 94230 Cachan, France}
\author[a]{Antonio Padilla}
\emailAdd{antonio.padilla@nottingham.ac.uk}

\date{\today}

\abstract{Vacuum energy sequestering is a mechanism for cancelling off the radiative corrections to vacuum energy. The most comprehensive low energy model, proposed in \cite{KP4}, has the ability to cancel off radiative corrections to vacuum energy for loops of both matter and gravitons, in contrast to earlier proposals. Here we explore the framework and cosmological consequences of this model, which we call   {\it Omnia Sequestra}, owing to its ability to sequester all. A computation of historic integrals on a cosmological background reveals some subtleties with UV sensitivity not seen in earlier models, but which  are tamed in a Universe that grows sufficiently old. For these old Universes, we  estimate the size of the radiatively stable residual cosmological constant and show that it does not exceed the critical density of our Universe today. We  also study the effect of phase transitions, both homogeneous and inhomogeneous, and find that generically spacetime regions with a small cosmological constant do not need to be fine-tuned against the scale of the transition, a result which is now seen to hold across all models of sequestering. The model is developed in other ways, from its compatibility with inflation, to the correct inclusion of boundaries and the  geometric consequences of certain choices of boundary data.}

\begin{document}

\maketitle

\section{Introduction}
In an effective quantum field theory, radiative corrections to the vacuum energy  generically scale like the fourth power of the cut-off. In the absence of gravity, this sensitivity to the ultra-violet (UV) sector of the theory is unobservable. However, once gravity is switched on, vacuum energy couples to the metric through the covariant measure, sourcing Einstein's field equations. The implications for the resulting spacetime  are catastrophic, curving the geometry to scales that lie at least sixty orders\footnote{Sixty orders of magnitude corresponds to a theory with a TeV cut-off, yielding a curvature scale of the order $(\textrm{TeV})^2/M_\text{Pl}$, sixty orders of magnitude above the Hubble scale, $H_0 \sim 10^{-33}$ eV.} of magnitude beyond the observational limits.  We can reconcile theory with experiment by including a finite correction to the counter term for vacuum energy although its value must be fine-tuned to one part in $10^{60}$ or worse.  Furthermore, this tuning is unstable against subtle changes to the UV sector of the theory, representing the worst  violation of naturalness known to Physics. This is known as  the cosmological constant problem \citecc.

Vacuum energy sequestering has  been proposed as a mechanism  for stabilising the observed cosmological constant against these radiative corrections to vacuum energy \citeseq. Sequestering exploits new rigid degrees of freedom, constant in space and time, whose global variation forces a cancellation between radiative corrections and a dynamical counterterm. The physics of the mechanism is strongly reminiscent of so-called decapitation \cite{decap,selftun}, with non-trivial behaviour kicking in at infinite wavelength.  The observed cosmological constant  turns out to be stable against radiative corrections, depending only on spacetime averages of local matter excitations, as well as boundary fluxes whose values are UV insensitive and should be fixed empirically.

Recently, sequestering has been shown to emerge at low energies from a pair of field theory monodromies \cite{mon} (see also \cite{irat})  raising hope for a realisation of the mechanism within string theory.
Although these monodromy constructions introduce new local degrees of freedom, they are taken to be extremely heavy, at the GUT scale or beyond, so that  at low energies we are only sensitive to their rigid behaviour. Such low energy descriptions coincide with the global modifications of General Relativity that characterise the original sequestering proposals \cite{KP1,KP2,KP3,KPSZ, KPS, KP4,etude}.  The most sophisticated version of these corresponds to what we shall call {\it Omnia Sequestra} (OS). This is the theory developed in \cite{KP4}, which has the capacity to cancel off radiative corrections to vacuum energy from loops that include virtual gravitons as well as matter fields.  The main goal of this paper is to elaborate further on this proposal, with particular emphasis on cosmological implications. 

Our first goal is to estimate the size of the residual cosmological constant in the theory. This is insensitive to radiative corrections to the vacuum energy, depending instead on historic integrals of localised matter sources, performed over the entire spacetime.  These integrals  reveal the presence of a potentially dangerous power law divergence,  coming from standard matter profiles  near  the cosmological singularity.  Such historic integrals, performed over the whole of spacetime, are required to calculate the spacetime averages and are an artefact of the global modification of gravity. Divergences were also present in earlier sequestering proposals \cite{KP2}, although in that case they were at most logarithmic, at least for homogeneous matter satisfying dominant energy conditions.  Of course, divergences themselves are not the issue since we do not expect our effective theory to apply to arbitrarily high scales. What matters is  how they scale with the cut-off. If we are to retain naturalness, {\it power law} scaling with the cutoff must not be allowed to contaminate the observed cosmological constant. Fortunately, in OS, the contamination can be diluted in a large and old universe, thanks to the spacetime averaging. In an infinite Universe there is no contamination whatsoever! In the end, we find that the historic integrals only make a small contribution to the residual cosmological constant, less than the current dark energy scale.  This was also the case in earlier models of sequestering \cite{KP2}, although the result is less trivial in this instance.

A significant part of our analysis will include a study of vacuum energy phase transitions in OS, both homogeneous and inhomogeneous,  the latter mirroring a similar analysis carried out for earlier versions of the sequestering proposal \cite{KPS}.  Although radiative (in)stability is  at the heart of the cosmological constant problem, phase transitions provide an added complication. This is because they  change the finite part of the vacuum energy, generically   by ${\cal O}(M^4)$, where $M$ is the scale of the  phase transition. For the QCD phase transition, we have $M \sim $ GeV, meaning that the vacuum energy before and after the transition differs by $ {\cal O}(\textrm{GeV}^4)$, which is some $46$ orders of magnitude larger than the critical density today, $ {\cal O}(\textrm{meV}^4)$.  Somehow, the Universe knew what energy  density to pick for the QCD vacuum prior to transition to an accuracy of one part in $10^{46}$, such that it would be cancelled to a sufficient extent after the transition. Electroweak and GUT phase transitions pose similar problems. 

The geometric response to phase transitions is therefore a crucial  question for sequestering models.  If the vacuum energy remains constant in time, it is successfully sequestered away, but when there is a jump, it is not immediately clear which value of the vacuum energy should get cancelled. It can certainly not be both since the dynamical counterterm is single valued across the entire spacetime\footnote{See, however,  \cite{lucas} for an interesting attempt to apply the mechanics of sequestering locally.}. In \cite{KP2, KPS} it was shown that for early sequestering proposals, vacuum energy is cancelled most efficiently in those regions of spacetime that dominate the four-volume. For homogeneous transitions, this means that the earlier the transition, the more efficiently it is sequestered at late times.  In an infinite Universe, sequestering is perfectly efficient after the last transition.  For inhomogeneous transitions mediated through bubble nucleation, one finds that, generically, it is the near-Minkowski vacua that dominate the four-volumes for the allowed configurations. This means no fine-tuning is required to achieve a near Minkowski solution, even in the presence of multiple vacua. 

 Qualitatively, we find that these results carry through to OS, although there are some differences in the detail.  For example,  for homogeneous transitions, the Universe needs to survive for a sufficiently long time, beyond the current epoch, to ensure that the jump in vacuum energy does not contaminate the late time value of the cosmological constant. We also find new constraints on the functional form of the scalar potentials in  the theory.  Despite these subtleties, our results do seem to indicate that the sequestering mechanism has the capacity to deal efficiently with phase transitions, without fine-tuning, regardless of which particular proposal one is interested in.

The rest of this paper is organised as follows: in the next section we shall review the OS proposal, which has the capacity to tame the effects of vacuum energy loops including virtual gravitons as well as matter fields. Our analysis will include a few additional nice-ities including an extension of the theory to include a spacetime boundary and an appropriate choice of boundary condition.  In section \ref{sec:homog}, we shall study  cosmological implications of OS, focussing on homogeneous configurations.  We perform estimates for spacetime averages by computing historic integrals, as well as studying the effects of homogeneous phase transitions and inflation. In section \ref{sec:bubble} we consider inhomogeneous tunnelling events mediated through bubble nucleation.  We compute tunnelling rates using Euclidian methods before Wick rotating back to Lorentzian signature to get an understanding of the geometric response seen by local observers.   As we stated above, sequestering cancellations are generically most efficient in regions which dominate the four-volume, favouring near Minkowski vacua without any need to fine-tune.  Finally, in section \ref{sec:conc}, we conclude.

\section{Omnia Sequestra} \label{sec:review}
The OS action is given by \cite{KP4},
\begin{multline}
S = \int \text{d}^4x \sqrt{-g} \left[ \frac{M_\text{Pl}^2}{2} R  - \Lambda (x) + \theta (x) R_{\text{GB}}  \right. \\ \left.  +\frac{1}{4!} \frac{\tensor{\epsilon}{^\mu^\nu^\lambda^\sigma}}{\sqrt{-g}} \left( \sigma \left(\frac{\Lambda }{\mu ^4}\right) \tensor{F}{_\mu_\nu_\lambda_\sigma} + \hat\sigma(\theta) \tensor{\hat F}{_\mu_\nu_\lambda_\sigma}\right)  \right]+S_\text{m}\left[\tensor{g}{^\mu^\nu},\Psi\right] \qquad \label{GS}
\end{multline}
where the metric $\tensor{g}{_\mu_\nu}$ has Ricci scalar $R$, and $R_{\text{GB}}=\tensor{R}{_\mu_\nu_\lambda_\sigma}\tensor{R}{^\mu^\nu^\lambda^\sigma} -4 \tensor{R}{_\mu_\nu}\tensor{R}{^\mu^\nu}+R^2$ is associated with the Gauss-Bonnet invariant.  The action also contains a pair of 4-form field stengths $\tensor{F}{_\mu_\nu_\lambda_\sigma} = 4\tensor{\partial}{_[_\mu}\tensor{A}{_\nu_\lambda_\sigma_]}$ and $\tensor{\hat F}{_\mu_\nu_\lambda_\sigma} = 4\tensor{\partial}{_[_\mu}\tensor{\hat A}{_\nu_\lambda_\sigma_]}$ each of whom is conjugate to the potential for a scalar field, respectively $ \sigma \left(\frac{\Lambda }{\mu ^4}\right)$ and $\hat\sigma(\theta)$. The scalars themselves correspond to a dynamical counterterm for the cosmological constant, $\Lambda(x)$,  and a linear  Gauss-Bonnet coupling,  $\theta(x)$.  At this stage, the potentials are assumed to be smooth functions with the argument of  $\sigma$  normalised by the scale,   $\mu$, generically assumed to lie at the cutoff. As for earlier versions of sequestering, both functions cannot be linear\footnote{Choosing both functions to be linear    would mean there was no way to  map  the  relevant observable (in this case,  the observed curvature on the  largest scales) to boundary data for either $\Lambda$ or $\theta$. }. $S_\text{m}$ describes the action for the matter fields minimally coupled to the metric. 

Note that the second integral in (\ref{GS}) is a non-gravitating, topological sector, as can be seen from the absence of the metric. The 4-forms act as a covariant measure and their gauge symmetries completely remove the local degrees of freedom, which fixes $\theta(x)$ and $\Lambda(x)$ on-shell. However, off-shell these scalars are fields, and their variation and selection of background values via the field equations constrain the spacetime average of the Gauss-Bonnet term, $\langle R_\text{GB} \rangle$. The couplings of $\theta(x)$ and $\Lambda(x)$ to the gravitational sector as well as to the 4-forms ensure the constraint on $\langle R_\text{GB} \rangle$, which yields the equation for the bare cosmological constant counterterm that guarantees cancellation of the loop corrections.

To illustrate, we will look at the global limit of the theory, and integrate out the 3-forms in \eqref{GS}. This constrains the scalars $\Lambda$ and $\theta$ to be constant in space and time. In other words, they become rigid degrees of freedom with no local variation, even off-shell. The resulting  effective action is given by,
\begin{eqnarray}
S = &&\int \text{d}^4x \sqrt{-g} \left( \frac{M_{\text{Pl}}^2}{2} R-\Lambda  + \theta R_{\text{GB}} \right)  +S_\text{m}\left[\tensor{g}{^\mu^\nu},\Psi\right] + 
 \sigma \left(\frac{\Lambda }{\mu ^4}\right) c + \hat\sigma(\theta) \hat c  \label{GSEA}
\end{eqnarray}
where the two scalars now behave as rigid Lagrange multipliers forcing global constraints on the theory. Here the constants $c=\int_\Sigma A$ and $\hat c=\int_\Sigma \hat A$ can be identified with  fluxes of the 3-forms $A=\frac{1}{3!}\tensor{A}{_\nu_\lambda_\sigma}dx^\nu \wedge dx^\lambda \wedge dx^\sigma$ and $\hat A=\frac{1}{3!}\tensor{\hat A}{_\nu_\lambda_\sigma}dx^\nu \wedge dx^\lambda \wedge dx^\sigma$ through the spacetime boundary, $\Sigma$.  Varying with respect to the metric and the rigid scalars, the resulting field equations are,
\be
M_\text{Pl}^2 G_{\mu\nu}= T_{\mu\nu}-\Lambda g_{\mu\nu}\label{GSFE1}
\ee
\be
 \frac{\sigma'}   {\mu^4}c=\int \sqrt{-g} \text{d}^4x,
 \qquad \hat\sigma' \hat c=-\int R_\text{GB} \sqrt{-g} \text{d}^4x 
 \label{GSFE2}
\ee
where $G_{\mu\nu}$ is the Einstein tensor and $T_{\mu\nu}=-\frac{2}{\sqrt{-g}} \frac{\delta S_m}{\delta g^{\mu\nu}}$ is the energy momentum tensor. The ratio of the latter two equations constrains the spacetime average of the Gauss-Bonnet term in terms of the boundary fluxes,
\be
\langle R_\text{GB} \rangle \myeq \frac{\int R_\text{GB} \sqrt{-g} \text{d}^4x }{\int \sqrt{-g} \text{d}^4x}=-\mu^4 \frac{\hat\sigma' \hat c}{  \sigma' c} \label{con}
\ee
This geometric constraint is crucial. The important point is that it is not scale invariant and so it constrains the infinite wavelength mode of the scalar curvature, which is the observable associated with the cosmological constant. Further, as $\hat c \to 0$, the action \eqref{GSEA} possesses a shift symmetry in the Gauss-Bonnet coupling, $\theta$, that protects the form of the constraint under graviton loops. In contrast, in earlier versions of sequestering there is a geometric constraint on the spacetime average of the Ricci scalar \cite{KP1, KP2, KPSZ} but this is   spoiled by graviton loops because there is no analogue of this shift symmetry for the Planck mass. In any event, for OS, it is the stability of the constraint \eqref{con} against graviton loops that allows us to extend the sequestering mechanism to take care of radiative corrections to vacuum energy that includes graviton loops as well as  matter  loops \cite{KP4}.

To see the cancellation of vacuum energy loops, we derive an effective gravity equation by eliminating the rigid degrees of freedom \cite{KP4},
\be \label{AvgFE}
M_\text{Pl}^2 {G}_{\mu\nu}={T}_{\mu\nu}-\frac{1}{4} \langle T \rangle g_{\mu\nu}- \Delta \Lambda g_{\mu\nu},
\ee
where $T=g^{\mu\nu} T_{\mu\nu}$ and $\Delta \Lambda$ is given by,
\be\label{resCC}
\Delta\Lambda^2=\frac{3M_\text{Pl}^4}{8}\Bigg[\left\langle R_{GB} \right\rangle - \left\langle (\tensor{W}{_\mu_\nu_\alpha_\beta})^2 \right\rangle 
  \left.
+\frac{2}{M_\text{Pl}^4} \left\langle \left({T}_{\mu\nu}-\frac{1}{4}T{g}_{\mu\nu}\right)^2 \right\rangle 
-\frac{1}{6 M_\text{Pl}^4}\left(\left\langle T^2 \right\rangle - \left\langle T \right\rangle^2\right) \right]
\ee
where $\tensor{W}{_\mu_\nu_\alpha_\beta}$ is the Weyl tensor. Now consider the effect of radiative corrections to vacuum energy. Firstly, note that  the regularized vacuum energy contributions to the energy momentum tensor, $\langle vac|\tensor{T}{^\mu_\nu}| vac \rangle = -V_\text{vac} \tensor{\delta}{^\mu_\nu}$,  will drop out of  terms on the second line of (\ref{AvgFE}).  The Weyl tensor is scale invariant so it too is immune from vacuum energy, while $\langle R_\text{GB} \rangle$ is constrained by \eqref{con}.  The latter contains a potential source of instability through its dependence on $\Lambda$ and $\theta$, which  receive radiative corrections that go as $\delta \Lambda \sim {\cal O}(M^4)$ and $\delta \theta \sim {\cal O}(1)\log(M/m)$, where $M$ is the effective field  theory cut-off and $m$ is a typical mass scale \cite{Demers}. However, as long as $\sigma$ and $\hat \sigma$ are smooth functions and $\mu\gtrsim M$, it is easy to see that radiative corrections to $\Delta \Lambda$ are no worse than an order one rescaling, $\delta \Delta \Lambda \sim {\cal O}(1) \Delta \Lambda$, in accordance  with naturalness.  Therefore, in OS, the conclusion is that the observed cosmological  constant is  stable against all radiative corrections to vacuum energy, including in the contributions  of virtual gravitons.

A more complete definition of OS should include boundary conditions and any  additional boundary terms required for a well defined variational principle, the analogue of the Gibbons-Hawking term in General Relativity
 \cite{GibbonsHawking}.  For an action of the form \eqref{GS}, the analogue of the Gibbons-Hawking boundary term is given by \cite{Davis, Boundary_terms},
\be \label{boundaryterm}
-\int_\Sigma d^3 x \sqrt{|h|} \left[ M_\text{Pl}^2 K+4 \theta (J-2 \hat G^{ij} K_{ij} )\right]
\ee
where $h_{ij}$ is the induced metric on the spacetime  boundary, $\Sigma$, with corresponding Einstein tensor $\hat G^{ij}$. The extrinsic curvature, $K_{ij} =-\frac12 {\cal L}_n h_{ij}$,  is defined in terms of the Lie derivative of the induced metric with respect to the {\it outward} pointing normal, $n^a$, and $K=h^{ij}K_{ij}$ is its trace. Finally, we define \cite{Davis},
\be
J_{ij}=\frac13 \left[(K_{kl}K^{kl} -K^2)K_{ij}+2K K_{ik} K^k{}_j-2K_{ik}K^{kl}K_{lj}\right]
\ee
along with its trace $J=h^{ij} J_{ij}$. 
The full action is now given by  \eqref{GS} supplemented with the boundary term \eqref{boundaryterm}). Its variation now yields a boundary contribution of the form  \cite{Davis, Boundary_terms},
\be
-\frac12 \int_{\Sigma }d^3 x \sqrt{|h|}\left[ \tensor{I}{^i^j}\delta \tensor{h}{_i_j} + I^{\theta}\delta \theta \right]
\ee
with,
\begin{eqnarray}
I^{ij} &&=-M_\text{Pl}^2 (K^{ij}-K h^{ij} ) -4\theta (3 J^{ij}-J h^{ij}+ 2 \tensor{\hat P}{^i^k^l^j}\tensor{K}{_k_l} )+\ldots \qquad  \label{Iij} \\
I^\theta &&= 8  (J-2 \hat G^{ij} K_{ij} ) \label{Ith}
\end{eqnarray}
where $\tensor{\hat P}{^i^k^l^j}$ is  the double dual of the Riemann tensor and the ellipsis denote terms proportional to gradients of $\theta$ that will vanish automatically thanks to the bulk equations of motion.

If we were to impose Dirichlet boundary conditions on all fields, the action and variational principle would now be well defined. However, as explained analogously in \cite{KPS},  Dirichlet boundary conditions on either $\theta$ or $\Lambda$ would suppress their off-shell  global fluctuations which are crucial to the success of the  sequestering mechanism.  To preserve the vacuum energy cancellation we must impose Neumann boundary conditions instead, 
\be
n^a \partial_a \delta \Lambda|_\Sigma=0, \quad n^a \partial_a \delta \theta|_\Sigma=0
\ee
Further imposing Dirichlet boundary conditions on the metric would now be problematic. Instead, we seek a boundary condition of the form $\delta h_{ij}|_\Sigma =A_{ij}\delta \theta|_\Sigma $ where $A_{ij}$ is chosen so that,
\be 
\left(\tensor{I}{^i^j}\delta \tensor{h}{_i_j} + I^{\theta}\delta \theta \right)|_\Sigma= 0  \, ,
\ee
guaranteeing a stationary action on-shell.  The task of finding a suitable choice of $A_{ij}$ is simplified for a three dimensional boundary by noting that the double dual of the Riemann tensor,  $\tensor{\hat P}{^i^k^l^j}$, vanishes identically in 3 dimensions.  We can also use the Cayley-Hamilton theorem  for a  $3 \times 3$ matrix, applied to $K_{ij}$, to show that $J_{ij}$ is a pure trace, $J_{ij}=-\frac23 h_{ij}  \det K$.  As a result, the final expression for \eqref{Iij} simplifies considerably, giving $I^{ij}=-M_\text{Pl}^2 (K^{ij}-K h^{ij} ) $. In the end, we found a one parameter ($z$) family of suitable choices for $A_{ij}$, 
\begin{multline}
A_{ij}=\frac{1}{M_\text{Pl}^2} \Bigg[ -16 \left(\hat R_{ij}-\frac14  \hat R h_{ij} \right)
-\frac{16}{3} \left( K_{ik} K^k{}_j- K K_{ij} -\frac{1}{4}(K_{kl} K^{kl}-K^2) h_{ij}   \right) \Bigg] \\
 +z \left[2 K K_{ij}+(K_{kl} K^{kl}-K^2 ) h_{ij} \right] \label{Aij}
\end{multline}
 We have not been able to establish an intuitive  geometric interpretation of this choice, although we note that for $z=0$, the extrinsic curvature terms appear in combinations familiar to the bulk curvature tensor, via the Gauss-Codazzi equations.

\section{Cosmological Implications}  \label{sec:homog}
Before studying the cosmological dynamics in detail, it is convenient to rewrite our effective gravity equation \eqref{AvgFE}  after  explicitly splitting  the energy-momentum tensor up into its constant vacuum energy part, $V_\text{vac}$ and local excitations,  described by $\tau_{\mu\nu}$.  To this end we write $T_{\mu\nu}=-V_\text{vac} g_{\mu\nu}+\tau_{\mu\nu}$ so that the vacuum energy drops out altogether and we  obtain,
\be
M_\text{Pl}^2 {G}_{\mu\nu}={\tau}_{\mu\nu}-\Lambda_\text{res} g_{\mu\nu}
\ee
where  we have a  residual cosmological constant given by,
\be \label{Lres}
\Lambda_\text{res}=\frac14 \langle \tau \rangle+\Delta \Lambda
\ee
with,
\begin{multline} \label{resCC2}
\Delta\Lambda^2=\frac{3M_\text{Pl}^4}{8}\Bigg[\left\langle R_{GB} \right\rangle - \left\langle (\tensor{W}{_\mu_\nu_\alpha_\beta})^2 \right\rangle 
  \left.
+\frac{2}{M_\text{Pl}^4} \left\langle \left({\tau}_{\mu\nu}-\frac{1}{4}\tau{g}_{\mu\nu}\right)^2 \right\rangle 
-\frac{1}{6 M_\text{Pl}^4}\left(\left\langle \tau^2 \right\rangle - \left\langle \tau \right\rangle^2\right) \right]
\end{multline}
where $\tau=g^{\mu\nu} \tau_{\mu\nu}$. As emphasized previously, this residual cosmological constant is stable against radiative corrections to the  vacuum energy and should now be fixed empirically. Of course, this is the same approach one takes for any relevant operator in effective field theory. For example, the electron mass is radiatively stable thanks to chiral symmetry, but its value cannot be predicted in effective field theory and should be set by measurement.

Let us now  focus on a homogeneous and isotropic background, described by the standard cosmological metric,
\be \label{frw}
ds^2=-dt^2+a(t)^2 d{\bf x}^2_\kappa
\ee
where $a(t)$ is the scale factor at time $t$, and $ d{\bf x}^2_\kappa$ is the metric on unit sphere ($\kappa=1$), plane ($\kappa=0$) or hyperboloid ($\kappa=-1$).  Assuming that the local matter content of the Universe is described by a homogeneous energy density, $\rho$ and  pressure, $p=w\rho$, we find that the dynamics is described by a Friedmann equation,
\be \label{frw1}
H^2+\frac{\kappa}{a^2}=\frac{\rho+\Lambda_\text{res}}{3M_\text{Pl}^2} 
\ee
where $\Lambda_\text{res}=-\frac14 \langle (1-3w)\rho \rangle +\Delta \Lambda$ and,
\be
\Delta \Lambda=
\pm \sqrt{\frac{1}{2}\langle \rho^2(1+3w) \rangle
 + \frac{1}{16}\langle \rho(1-3w) \rangle^2- \frac{3}{8} M_\text{Pl}^4 \mu^4 \frac{\hat \sigma'}{\sigma'} \frac{\hat c}{c} } \label{DL}
\ee
Here we have used the fact that the Weyl tensor vanishes on a Friedmann-Robertson-Walker metric \eqref{frw}, as well as the constraint \eqref{con}.  Following \cite{KP2} , we  evaluate the  spacetime averages by assuming that the cosmology takes place over a (regulated) proper time interval $t_\text{in} < t < t_\text{out}$, with a  (regulated) spatial co-moving volume $\text{Vol}_3$.  For example, when we explicitly compute the constraint \eqref{con} in this way we obtain,
\be
\langle R_{GB} \rangle \myeq \frac{\int_{t_\text{in}}^{t_\text{out}} dt ~a^3 \left[24 \frac{\ddot a}{a} \left(H^2+\frac{\kappa}{a^2} \right)\right] }{\int_{t_\text{in}}^{t_\text{out}}dt ~a^3}=-\mu^4 \frac{\hat\sigma' \hat c}{  \sigma' c} \label{con1}
\ee
 The cancellation of the  spatial volumes will be generic for all spacetime averages computed on this background.

\subsection*{Calculation of historic integrals}

Let us now estimate the historic integrals that appear in \eqref{Lres} and \eqref{resCC2}. To  do so, we follow \cite{KP2} and split the cosmological history into intervals $(t_i,t_{i+1})$,  for which the dominant  source, $\rho$,  has  equations of state $w_i$ and the cosmological evolution has an {\it effective} equation of state $\bar w_i$. Generically we expect $w_i=\bar w_i$, although exceptions could include an epoc of curvature domination or domination by the residual cosmological constant, $\Lambda_\text{res}$, as one might expect to see at late times.  In this $i$th interval, we can use the energy conservation equation $\dot \rho=-3H(\rho+p)$ and the Friedmann equations \eqref{frw1} to  obtain,
\be \label{FEqs}
H=H_{i+1}\left(\frac{a}{a_{i+1}}\right)^{-\frac{3}{2}\left(1+\bar w_i\right)}, \, \rho=\rho_{i+1}\left(\frac{a}{a_{i+1}}\right)^{-3\left(1+ w_i\right)},
\ee
where $a_j$ and $H_j$ denote the scale and Hubble factors evaluated at time $t_j$. Let us define the generic contributions to the integrals in \eqref{DL} and evaluate them using \eqref{FEqs}. For $n=0, 1, 2$ we write,
\begin{eqnarray}
I_{n, i} &\myeq& f_{n, i}  \int_{t_i}^{t_{i+1}} dt \,a^3 \rho^n \\
&=& \left(\frac{a^3 \rho^n}{H}\right)_{i+1} \frac{ f_{n, i}}{g_{n, i}} \left[1-\left(\frac{a_i}{a_{i+1}}\right)^{g_{n, i}} \right] \\
&=& \left(\frac{a^3 \rho^n}{H}\right)_{i} \frac{ f_{n, i}}{g_{n, i}}\left[\left(\frac{a_{i+1}}{a_{i}}\right)^{g_{n, i}}-1 \right] 
\end{eqnarray}
where,
\be
f_{n, i}=\begin{cases} 1 & n=0 \\
1-3w_i & n=1 \\
1+3w_i & n=2
\end{cases}
\ee
and,
\be
 g_{n,i}=\frac32 (3+\bar w_i)-3n(1+ w_i)
 \ee Note that for $g_{n, i}=0$, we understand the formulae for $I_{n, i}$ by taking the limit as $g_{n, i}\to 0$, in which case we obtain logarithms. Let us  also define $I_n=\sum_i I_{n, i}$ where the sum is performed over all intervals in the entire cosmic history, so that now we may write,
\be \label{Lres1}
\Lambda_\text{res}=-\frac14 \frac{I_1}{I_0}\pm \sqrt{\frac12 \frac{I_2}{I_0}+\frac{1}{16}\left(  \frac{I_1}{I_0}\right)^2 -\frac{3}{8} M_\text{Pl}^4  \mu^4 \frac{\hat \sigma'}{\sigma'} \frac{\hat c}{c}}
\ee
Owing to the quadratic nature of the global constraint, our solution comes with two roots. At this stage, we have no compelling reason to pick one root over the other. In higher dimensional Gauss-Bonnet gravity, solutions also split into two branches, and it is the branch that admits a smooth Einstein limit that typically avoids pathological behaviour \cite{GB}.

Consider first an expanding phase, so that adjacent intervals satisfy $a_{i-1} \ll a_i \ll a_{i+1}$.  We obtain the following ratio, 
\be
\left|\frac{I_{n, i}}{I_{n, i-1}} \right|=\frac{\frac{ f_{n, i}}{g_{n, i}}\left[\left(\frac{a_{i+1}}{a_{i}}\right)^{g_{n, i}}-1 \right]  }{\frac{ f_{n, i-1}}{g_{n, i-1}} \left[1-\left(\frac{a_{i-1}}{a_{i}}\right)^{g_{n, i-1}} \right]}
\ee
Depending on the values for the $g_{n, i}$, there are three possible scenarios\footnote{When $g_{n,i}>0, g_{n,i-1} <0$ we could in principle be in any of the three cases, depending on the relative size of the scale factors.}:
\begin{enumerate}
\item  $|I_{n, i} |\gg |I_{n, i-1}|$ e.g. when $g_{n,i}>0, g_{n,i-1} >0$,   \label{ir}
\item  $|I_{n, i} |\sim |I_{n, i-1}|$ ~e.g. when $g_{n,i}<0, g_{n,i-1} >0$,
\item  $|I_{n, i} |\ll |I_{n, i-1}|$ e.g. when $g_{n,i}<0, g_{n,i-1} <0$\label{uv}
\end{enumerate}
When case \ref{ir} occurs, the later interval dominates thanks to the largeness of $a_{i+1}$, for appropriate values of $g_{n, i}$. In contrast, when case \ref{uv} occurs, the earlier interval dominates thanks to the smallness of $a_{i-1}$, again, for appropriate values of $g_{n, i}$. One can obtain analogous results in a contracting phase (if there is one). What all this tells us is that the sums $I_n$ are dominated by their extreme infra-red and ultra-violet intervals, where the scale factor is largest and smallest respectively.  To develop this further,  let us define the infra-red interval as $a_\star< a< a_\text{max}$ and the ultra-violet interval as $a_\text{min}<a < a_\dagger$, where $a_\text{max}$ is the largest scale factor in the cosmic history, and $a_\text{min}$ is the smallest.  $a_\text{min}$  is not taken to be zero, as one might naively expect, but to a regulated finite value consistent with the UV  cut-off of the theory.  In contrast, we do allow $a_\text{max}$ to be infinite, in principle.  The precise values of $a_\star$ and $a_\dagger$ are not important in what follows. We may now write\footnote{In an expanding then contracting Universe, we would get UV and IR contributions from both phases, but we suppress this sum in the interests of brevity.},
\be
I_n \sim I_n^{UV}+I_n^{IR}
\ee
where,
\begin{eqnarray}
I_n^{UV} &=&\left(\frac{a^3 \rho^n}{H}\right)_{\star} \frac{ f_{n, UV}}{g_{n, UV}} \left[1-\left(\frac{a_\text{min}}{a_{\star}}\right)^{g_{n, UV}}\right] \\
I_n^{IR} &=&\left(\frac{a^3 \rho^n}{H}\right)_{\dagger} \frac{ f_{n, IR}}{g_{n, IR}} \left[\left(\frac{a_\text{max}}{a_{\dagger}}\right)^{g_{n, IR}}-1\right] 
\end{eqnarray}
These terms contain possible divergences as $a_\text{min} \to 0$ (for $g_{n, UV} \leq 0$) and $a_\text{max} \to \infty$ (for $g_{n, IR} \geq 0$). Of course, what we are really interested in are the ratios $I_n/I_0$. To this end we note that $g_{0, i}=\frac32(3+\bar w_i) \in [3, 6]$ for an effective equation of state $\bar w_i \in [-1, 1]$. This range is consistent with  sources that satisfy the dominant energy condition. In any event, it follows that there is no divergent UV contribution to $I_0$, so that we simply have,
\be \label{I0}
I_0 \sim I_0^{IR}  \sim \left(\frac{a^3}{H}\right)_{\dagger} \frac{1}{\frac32(3+\bar w_{IR} )} \left(\frac{a_\text{max}}{a_{\dagger}}\right)^{\frac32(3+\bar w_{IR})}
\ee
Now consider the ratios. From the infra-red regime, we have,
\begin{eqnarray}
\frac{I_n^{IR}}{I_0} &\sim&  \frac32(3+\bar w_{IR} )  \rho^n_\dagger \frac{ f_{n, IR}}{g_{n, IR}}  \left(\frac{a_\text{max}}{a_{\dagger}}\right)^{-3n(1+ w_{IR})} \nonumber \\
& \sim&  \frac32(3+\bar w_{IR} )  \rho^n_\text{max} \frac{ f_{n, IR}}{g_{n, IR}}
\end{eqnarray}
The matter equation of state satisfies the dominant energy condition with vacuum energy excluded\footnote{The constant underlying  vacuum energy gets sequestered. We will deal wth vacuum energy phase transitions in the next section.}, $w_i \in(-1, 1]$, so there is no divergence in the ratio $I_n^{IR}/I_0$ for $n=1, 2$. Indeed, we see that this ratio scales like $ \rho^n_\text{max}$, where $\rho_\text{max}$ is the homogeneous energy density associated with localised matter sources at the point where the Universe is at its largest. This contribution vanishes in an infinite Universe thanks to the dilution of such sources. 

Now consider the ultra-violet regime. Here we have,
\be
\frac{I_n^{UV}}{I_0} \sim - \frac32(3+\bar w_{IR} ) \frac{\left(\frac{a^3 \rho^n}{H}\right)_{\star}}{\left(\frac{a^3}{H}\right)_{\dagger}}  \frac{ f_{n, UV}}{g_{n, UV}} \frac{\left[\left(\frac{a_\text{min}}{a_{\star}}\right)^{g_{n, UV}}-1\right]}{ \left(\frac{a_\text{max}}{a_{\dagger}}\right)^{\frac32(3+\bar w_{IR})} }
\ee
For $g_{n, UV} <0$, there is a dangerous {\it power law} divergence as $a_\text{min} \to 0$ in a finite Universe (where $a_\text{max}$ is finite). Such a divergence could contaminate the observed cosmological constant, $\Lambda_\text{res}$, with power law cut-off dependence, in violation of naturalness.  Indeed, given the allowed values $w_i \in(-1, 1], \bar w_i \in [-1, 1]$, we have that $g_{n, i} \in [3-6n, 6)$ and therefore a potentially dangerous cut-off dependence for $n=1, 2$. If we choose to identify $\bar w_{UV}=w_{UV}$, we can reduce the cut-off scaling to at worst a logarithmic one (for $w_{UV}=1$) for $n=1$ \cite{KP2}, although for $n=2$, power law dependence remains for $w_{UV} \in [-1/3, 1]$. 

This unnatural cut-off dependence can be eliminated in an infinite Universe, thanks to the volume suppression as $a_\text{max} \to \infty$.  This suggests that there is a lower bound on the size of the Universe set by naturalness.  Let's have some fun by estimating this, noting first that $\frac{I_n^{UV}}{I_0} \sim  \left(\frac{a^3 \rho^n}{H}\right)_{\text{min}}/\left(\frac{a^3}{H}\right)_{\text{max}} $. If we take $\rho_\text{min} \sim M_\text{Pl}^2 H^2_\text{min}$ then we can write,
\begin{eqnarray}
\frac{I_n^{UV}}{I_0} &\sim& \left(\frac{\mathcal{N}}{a_\text{max}/a_0} \right)^3 \frac{H_\text{max}}{H_0} (M_\text{Pl}^2 H_0^2)^n  \\
&\lesssim& \left(\frac{\mathcal{N}}{a_\text{max}/a_0} \right)^3  (M_\text{Pl}^2 H_0^2)^n
\end{eqnarray}
where $a_0$ is the present day scale factor and,
\be
\mathcal{N}=\left(\frac{1}{H_0 l_{UV}}\right)^{\frac13 \left[2n-1-\frac{2}{1+\bar W(a_\text{min}, a_0)}\right]}
\ee
Here we have integrated over the cosmic history from the cutoff to the present day, giving,
\be
\frac{H_\text{min}}{H_0}= \left(\frac{a_0}{a_\text{min}} \right)^{\frac32(1+\bar W(a_\text{min}, a_0,))}
\ee
where,
\be
1+\bar W(a_\text{min}, a_0)=\frac{ \int^{\ln a_0}_{\ln a_\text{min}} d\ln a  ~(1+\bar w (\ln a)) }{{ \int^{\ln a_0}_{\ln a_\text{min}} d\ln a }}
\ee
 and $\bar w (\ln a)$ is the effective equation of state when the scale factor has size $a$.  We have also assumed  $H_\text{min}\sim l_{UV}^{-1}  $ where $l_{UV}$ is the length scale at which we cut off the theory (possibly the string length or the Planck length). In any event, provided $a_\text{max}/a_0 \gtrsim \mathcal{N}$, we are guaranteed that the UV contribution does not exceed the scale set by the critical density today, $I_n^{UV}/I_0 \lesssim (M_\text{Pl}^2 H_0^2)^n$.

The condition $a_\text{max}/a_0 \gtrsim \mathcal{N}$ is only required for  $n=1, 2$,  and given that $\bar W \in [-1, 1]$ our strongest bound comes from $n=2$ and $\bar W=1$. This yields $a_\text{max}/a_0   \gtrsim (H_0 l_{UV})^{-2/3}$, which for a Planckian cut-off, is  a comforting  $
a_\text{max}/a_0   \gtrsim 10^{40} 
$
or $92$ more efolds of expansion!  In any event, we trust that the reader has enough time to finish going through  the rest of this paper. 

Bringing everything together, we see that the residual cosmological constant receives  up to three distinct contributions:  the IR part of the historic integrals scaling as $\rho_\text{max} \lesssim M_\text{Pl}^2 H_0^2$; the UV part of the historic integrals scaling as  
$\left(\frac{\mathcal{N}}{a_\text{max}/a_0} \right)^\frac{3}{n} \left(\frac{H_\text{max}}{H_0}\right)^\frac{1}{n} M_\text{Pl}^2 H_0^2 \lesssim M_\text{Pl}^2 H_0^2$; and the flux contribution, scaling as 
 $\Lambda_\text{flux}=\sqrt{-\frac{3}{8} M_\text{Pl}^4  \mu^4 \frac{\hat \sigma'}{\sigma'} \frac{\hat c}{c}}$.  The latter can be fixed empirically and assumed to lie below the dark energy scale. In conclusion, then, provided the Universe grows sufficiently large, the residual cosmological constant will not exceed the critical density of the Universe today.

\subsection*{Homogeneous phase transitions}

We now consider the effect of a single homogeneous phase transition in the vacuum energy. As explained in the introduction, such transitions shift the  potential by a constant amount ${\cal O}(M^4)$, where $M$ is the scale of the transition, with well known examples being the electroweak and the QCD phase transitions.  Assuming a rapid transition, we  can model this by a step function of size  $\Delta V=V_2-V_1$, at time $t_*$, so that the energy momentum tensor is given by $\tensor{T}{_\mu_\nu}=-V(t)\tensor{g}{_\mu_\nu}+\tau_{\mu\nu}$, where,
\be \label{PT}
V(t) = \left\{
            \begin{array}{ll}
                V_1 & \quad t < t_* \\
                V_2 & \quad t > t_*
            \end{array}
        \right.
\ee
and $\tau_{\mu\nu}$ represents localised sources with equation of state in the range $(-1, 1]$, consistent with the dominant energy condition.  In what follows, we will make use of  the following shorthand for the spacetime volume before transition,
\be
\Omega_1=\text{Vol}_3 \int_{t_\text{in}}^{t_*} dt ~a^3\,
\ee
 the spacetime volume after,
\be
\Omega_2=\text{Vol}_3 \int^{t_\text{out}}_{t_*} dt ~a^3
\ee
and their ratio $\mathcal{I}=\frac{\Omega_2}{\Omega_1}$. We also define the  following ``before" and ``after" averages, respectively, 
\be
 \langle \tau\rangle_1=\frac{\text{Vol}_3 \int_{t_\text{in}}^{t_*} dt ~a^3 \tau}{\Omega_1}, \qquad  \langle \tau\rangle_2=\frac{\text{Vol}_3 \int^{t_\text{out}}_{t_*} dt ~a^3 \tau}{\Omega_2}
\ee
Finally we introduce the local excitation of the potential, 
 \be
 \delta V=V(t)-\langle V\rangle =\begin{cases}
 -\Delta V \frac{\mathcal{I}}{(1+\mathcal{I}) }& t<t_* \\
 \Delta V \frac{1}{(1+\mathcal{I}) } & t>t_*
 \end{cases}
 \ee
We are now ready to write down the  effective gravity equation in the presence of a homogeneous transition. It is given by
$
M_\text{Pl}^2  G_{\mu\nu}=-\Lambda_\text{eff}(t) g_{\mu\nu}+\tau_{\mu\nu}
$
where the effective cosmological constant is,
\be
\Lambda_\text{eff}(t)=\delta V +\Delta \Lambda+\frac14 \langle \tau \rangle
\ee
and,
\begin{multline}
\Delta \Lambda^2=
- \frac{\mathcal{I}}{(1+\mathcal{I})^2} \left[(\Delta V)^2 -\frac12 \Delta V \left( \langle \tau\rangle_2 -\langle \tau \rangle_1
\right)\right] \\
+\frac34\left\langle \left({\tau}_{\mu\nu}-\frac{1}{4}\tau{g}_{\mu\nu}\right)^2 \right\rangle -\frac{1}{16}\left(\left\langle \tau^2 \right\rangle - \left\langle \tau \right\rangle^2\right) 
  -\frac{3}{8} M_\text{Pl}^4  \mu^4 \frac{\hat \sigma'}{\sigma'} \frac{\hat c}{c} 
\end{multline}
For $\Delta V=0$, this result reduces to  \eqref{Lres} and \eqref{resCC2} for vanishing Weyl tensor, as of course it should. To study the effect of the phase transition, we focus on the $\Delta V$ dependent terms in our expression. These introduce some time dependence in the effective cosmological constant, through $\delta V$. To develop some intuitive understanding let us first consider very early and very late transitions. For a very early transition, we expect $\mathcal{I} \gg 1$ and so to get some insight we  take the limit $\mathcal{I} \to \infty$.  In this case, the effective cosmological constant after the transition loses all knowledge of the scale of the jump.   Prior to the transition, the effective cosmological constant is strongly sensitive to $\Delta V$.  In contrast, for late transitions, modelled intuitively with the limit $\mathcal{I} \to 0$, we have the opposite: no sensitivity to $\Delta V$ prior to transition, but strong sensitivity after.  Although the details are different, these conclusions are qualitatively the same as for earlier models of sequestering: sequestering works best in the volume that dominates the spacetime.  This means that we always have late time suppression of the jump  for early transitions \cite{KP2}.

  Let us now estimate the size of this volume  ratio and the impact on the effective cosmological constant more carefully. As we saw in the previous section, historic integrals are generically dominated by the period in which the Universe is largest.  This corresponds to the latest time during an expanding phase.  We shall consider phase transitions occurring in the past, during expansion, consistent with the structure of the Standard Model. The results of the previous section (see equation \eqref{I0} and use \eqref{FEqs})   then suggest that,
 \be
 \Omega_1+\Omega_2 =I_0 ={\cal O}(1)\left(\frac{a^3}{H} \right)_\text{max},  \quad \Omega_1 ={\cal O}(1) \left(\frac{a^3}{H} \right)_*
 \ee 
and so,
\be
\mathcal{I} ={\cal O}(1) \left(\frac{a_\text{max}}{a_*}\right)^3 \frac{H_*}{H_\text{max}} -1 \sim {\cal O}(1) \left(\frac{a_\text{max}}{a_*}\right)^3 \frac{H_*}{H_\text{max}}
\ee
where we have used the fact that  $a_\text{max} \gg a_*$ and $H_\text{max} \ll H_*$. Since $\mathcal{I} \gg 1$, we have that,
\be \label{DLjump}
\Delta \Lambda^2 \approx -\left[\frac{(\Delta V)^2}{\mathcal{I}} -\frac{ \Delta V}{2\mathcal{I}} \left( \langle \tau\rangle_2 -\langle \tau \rangle_1
\right)\right]+\ldots
\ee
where $\ldots$ denote transition independent terms and,
\be
\delta V \approx \begin{cases}
 -\Delta V &  t<t_* \\
  \frac{\Delta V}{\mathcal{I} } & t>t_* \end{cases}
\ee
As anticipated, we get strong dependence on the scale of the jump, prior to the transition. This will yield a short burst of inflation just before the transition occurs. After the transition, it would seem that any dependence on the scale of the jump is heavily suppressed. To see by how much, recall that  integrating the cosmic history from the transition to the maximum size, we can show that,
\be
\frac{H_*}{H_\text{max}}= \left(\frac{a_\text{max}}{a_*} \right)^{\frac32(1+\bar W(a_*, a_\text{max}))}
\ee
where $1+\bar W(a_*, a_\text{max})=\frac{ \int_{\ln a_*}^{\ln a_\text{max}} d\ln a  ~(1+\bar w (\ln a)) }{{ \int_{\ln a_*}^{\ln a_\text{max}} d\ln a }}$.  It then follows that the contribution to $\delta V$  after the transition  goes as,
\be
\delta V_\text{after}\approx   \frac{\Delta V}{\mathcal{I} }  = {\cal O}(1)   \frac{\Delta V}{M_\text{Pl}^2 H_*^2 } \left( \frac{H_\text{max}}{H_*}\right)^\frac{1-\bar W }{1+\bar W} M_\text{Pl}^2 H_\text{max}^2
\ee
We expect $|\Delta V| = {\cal O}(1)  M_\text{Pl}^2 H_*^2$ and so since $\bar W(a_*, a_\text{max}) \in [-1,1]$, it follows that this contribution is no larger than the critical density at maximum size,  or indeed the critical density today, $\delta V_\text{after} \lesssim  M_\text{Pl}^2 H_\text{max}^2  \lesssim M_\text{Pl}^2 H_0^2$. This reflects similar conclusions drawn in \cite{KP2}. In an infinitely old,  asymptotically de Sitter Universe,  we get  exponential suppression since $\bar W(a_*, a_\text{max})= -1$.

Now consider the jump contributions to $\Delta \Lambda$ as shown in \eqref{DLjump}. Similar considerations yield,
\be
\frac{(\Delta V)^2}{\mathcal{I} }  = {\cal O}(1) \left( \frac{\Delta V}{M_\text{Pl}^2 H_*^2 } \right)^2 \left( \frac{H_\text{max}}{H_*}\right)^{-\frac{1+3\bar W}{1+\bar W} } M_\text{Pl}^4 H_\text{max}^4
\ee
For the other contribution, we adapt the results of the previous section to estimate the ``before" and ``after" averages as $\langle \tau\rangle_1 \sim  {\cal O}(1)  \rho_*  \gg \langle \tau \rangle_2 \sim  {\cal O}(1)  \rho_\text{max}$. This then gives the scale,
\be
\frac{ \Delta V}{2\mathcal{I}} \left( \langle \tau\rangle_2 -\langle \tau \rangle_1
\right) = {\cal O}(1)   \frac{\Delta V}{M_\text{Pl}^2 H_*^2 }    \frac{\rho_*}{M_\text{Pl}^2 H_*^2 }   \left( \frac{H_\text{max}}{H_*}\right)^{-\frac{1+3\bar W}{1+\bar W} } M_\text{Pl}^4 H_\text{max}^4
\ee
Assuming $|\Delta V|, \rho_* = {\cal O}(1)  M_\text{Pl}^2 H_*^2 $ the  result is that the jump contributions to $\Delta \Lambda$ both come in at the scale,
\be
[\Delta\Lambda]_\text{jump}= {\cal O}(1)  \left( \frac{H_\text{max}}{H_*}\right)^{-\frac{1+3\bar W}{2(1+\bar W)} }  M_\text{Pl}^2 H_\text{max}^2
\ee
 In contrast to $\delta V_\text{after}$, this contribution has the potential to be enhanced relative to the critical density  at maximum size $M_\text{Pl}^2 H_\text{max}^2$, whenever $\bar W \in (-1/3, 1]$.   This enhancement could easily make $\Delta \Lambda$ larger than the critical density today.  Requiring that this is {\it not} the case imposes the following bound,
 \be
 \bar W <-\frac13\left(\frac{1-4r}{1-\frac43 r}\right), \qquad r=\frac{\ln \frac{H_0}{H_\text{max}}}{\ln \frac{H_*}{H_\text{max}}}
 \ee
 where we have assumed $r<\frac34$. As we have stated previously, in an infinitely old,  asymptotically de Sitter Universe,  we get   $\bar W(a_*, a_\text{max})= -1$ and so there are no dangerously large contributions to $\Delta \Lambda$.  But what if the current de Sitter phase is only transient?  Let's have more fun and estimate how  long this quasi de Sitter stage needs to last in order to ensure there is no dangerous enhancement of $[\Delta\Lambda]_\text{jump}$. To do this, we crudely model the history of the universe as radiation dominated from $a_*$ until $a_\text{eq}$, then matter dominated from $a_\text{eq}$ until $a_\text{de}$, and finally quasi-de Sitter behaviour from $a_\text{de}$ until $a_\text{max}$. We shall not assume that $a_\text{max}$ is infinite, allowing for the possibility that the quasi de Sitter stage comes to an end close to the maximum size. In any event, we find that,
 \be
 \bar W=\frac{\ln \left[\left(\frac{a_\text{eq}}{a_*}\right)^\frac43\frac{a_\text{de}}{a_\text{eq}}\right]}{\ln \frac{a_\text{max}}{a_*}}-1
 \ee
Assuming $r$ to be small then requiring $\bar W <-1/3$, we obtain the following lower bound on the would-be size of the Universe,
\be
a_\text{max}>\frac{\sqrt{a_\text{eq} a_\text{de}}}{a_*}a_\text{de}
\ee
To bring this to life, we note that the QCD phase transition, matter-radiation equality and matter-dark energy equality occur at redshifts of $10^{12}, 3400$ and $0.4$ respectively. Setting $a_* \sim a_{QCD}$, our bound then implies 
$
a_\text{max}/{a_0} \gtrsim 10^{10}
$
which is less constraining than our estimate in the previous section. Earlier transitions would suggest a longer future, of course.

\subsection*{Inflation}
We have seen in previous sections how a large and old Universe can eliminate potentially large and unnatural contributions to the residual cosmological constant.  The standard mechanism for achieving a large Universe is through inflation so it is natural to ask if it can be embedded in a theory of OS. We might be concerned that the inflaton source behaves like a constant vacuum energy to zeroth order in slow roll and will therefore be sequestered. This conclusion is too quick, however.  Inflation resembles a (slow) phase transition and,  as we have just seen, the corresponding scale is visible in the effective cosmological constant {\it prior} to the end of the transition. Compatibility with inflation was shown for  earlier models of sequestering \cite{KP2},  and we will now show that this is also the case here.

We assume, for simplicity, standard single field inflation (for a review, see \cite{Lindeinf}), described by a canonical scalar $\varphi$ with potential $V(\varphi)$, minimally coupled to the metric. During inflation,  all other sources of energy-momentum are quickly diluted away,  and,  during slow roll, we have that the effective Friedmann equation and energy conservation equation are  given by,
\be
H^2 \approx \frac{V+\Lambda_\text{res}}{3M_\text{Pl}^2}, \qquad 3H \dot \varphi \approx -V'
\ee
where  we have also neglected spatial curvature. We now ask whether or not the inflationary contribution to  the residual cosmological constant can significantly affect the dynamics. If inflation were to go on like this forever, the answer would be ``yes'', since the sequestering mechanism would force an exact cancellation between a constant value for $V$ and $\Lambda_\text{res}$. Of course, inflation must end, and it turns out that its contribution to  $\Lambda_\text{res}$ is nowhere near large enough to compete with the potential.

To see this, let us now estimate the inflationary contribution to $\Lambda_\text{res}$. Again, assuming slow roll, we have that $ \tau_{\mu\nu} \approx - V(\varphi)g_{\mu\nu}$. It follows that,
\be
\langle \tau \rangle \approx -4 \frac{ \int_{t_\text{start}}^{t_\text{end}}  dt ~a^3 V(\varphi)}{\int_{t_\text{in}}^{t_\text{out}} dt~a^3}
\ee
where inflation starts at time $t_\text{start} \approx t_\text{in}$ and ends at time $t_\text{end} \ll t_\text{out}$. We can estimate the integrals to give,
\be
\langle \tau \rangle  = {\cal O}(1) V_\text{inf}\left( \frac{a_\text{end}}{a_\text{max}}  \right)^3  \frac{H_\text{max}}{H_\text{inf}} 
\ee
where $V_\text{inf}=M_\text{Pl}^2 H_\text{inf}^2$ and $H_\text{inf}^2$ is the scale of inflation. Since $H_\text{max} \ll H_\text{inf}$ and $a_\text{end} \ll a_\text{max}$ we have that $|\langle \tau \rangle|$ is much less than the scale of the potential during inflation $V_\text{inf}$. Similarly, we find that, 
\be
\langle \tau^2 \rangle  = {\cal O}(1) V^2 _\text{inf}\left( \frac{a_\text{end}}{a_\text{max}}  \right)^3  \frac{H_\text{max}}{H_\text{inf}}  \ll V^2 _\text{inf}
\ee
and $\left \langle \left(\tau_{\mu\nu} -\frac14 \tau g_{\mu\nu} \right)^2 \right\rangle \approx 0$. Since the flux contribution, $\Lambda_\text{flux} \lesssim M_\text{Pl}^2 H_0^2 \ll V_\text{inf}$, we conclude that, $|\Lambda_\text{res}| \ll V_\text{inf}$, or in other words, inflation in OS goes through as normal.
\subsection*{Geometric consequences of choosing the flux}

The boundary fluxes, given by $c$ and $\hat c$,  are taken to be infra-red geometric quantities, whose values are simply given as  fixed boundary conditions in the effective field theory.  Nevertheless, it is interesting to explore the consequences of particular choices.  For example, in an homogeneous universe,  vanishing $\hat c$ forces the spatial curvature to be negative, consistent with  a spatially open Universe.  To see this we simply set $\hat c=0$ in \eqref{con1}, then solve the integral to give,
\be
\kappa|_{\hat c=0}=-\frac{[\dot a ^3]_{t_\text{in}}^{t_\text{out}}}{3[\dot a]_{t_\text{in}}^{t_\text{out}}}
\ee
The right hand side of this expression is negative for all real choices of $\dot a_\text{in}$ and $\dot a_\text{out}$.
We emphasize that for generic $\hat c$, there are no such well defined constraints on  the spatial geometry. Indeed, more generally we have from \eqref{con1},
\be
\kappa=-\frac{[\dot a ^3]_{t_\text{in}}^{t_\text{out}}}{3[\dot a]_{t_\text{in}}^{t_\text{out}}}-\mu^4 \frac{\hat\sigma' \hat c}{  \sigma' c}\frac{\int_{t_\text{in}}^{t_\text{out}} dt ~a^3}{24 [\dot a]_{t_\text{in}}^{t_\text{out}}}
\ee
where the second term can take either sign and be as large or small as we like, depending on the choices for the flux and the cosmological dynamics. 
\section{Inhomogeneous Phase Transitions}  \label{sec:bubble}
Transitions in vacuum energy can also occur locally through bubble nucleation.  In standard Einstein gravity, the formalism for describing this was pioneered by Coleman and  collaborators \cite{Coleman_1, Coleman_2, Coleman_DeLuccia} and adapted to early models of sequestering in \cite{KPS}. There it was shown that vacuum energy was most efficiently sequestered in regions of spacetime of largest volume, favouring near-Minkowski configurations without fine-tuning.  We shall now show that similar conclusions can be drawn for OS.

First we assume a potential that interpolates between two minima,  separated by a scale $\Delta V$. Tunnelling from one vacuum to the other can occur via spontaneous nucleation of a spherical bubble containing the new vacuum in the interior, then expanding at the speed of light.  As we will see, not all configurations are kinematically allowed, at least if we assume a sensible microscopic structure in the bubble wall. Further, for the kinematically allowed configurations, we can estimate the rate of transition per unit volume by computing the so-called bounce solution to the Euclidean field equations.

Let us proceed by first computing the bounce.  As usual, we  will work in the thin wall approximation \cite{Coleman_DeLuccia}, and assume that  the bounce solution is O(4) invariant \cite{O4_1,O4_2}. Under these assumptions we can write the metric with the ansatz $ds^{2} = dr^{2} + \rho^{2}(r) d\chi^{2}$ where $d \chi^{2}=\gamma_{ij}dx^idx^j$ is the unit 3-sphere. In a neighbourhood of the bubble wall, we adopt a coordinate system with the wall at $r=0$, the bubble exterior corresponding to $r>0$ (which we will call denote $\mathcal{M_+}$), and the interior $r<0$ (which we will denote $\mathcal{M_-}$).  We shall also refer to the exterior as the ``old" vacuum, and to the interior as the ``new".  The rotational invariance allows us to write all fields as functions of the radial coordinate $r$ only. For example, the $3$-forms components are now,
\be
A_{ijk}=A(r)\sqrt{\gamma}\epsilon_{ijk}, \qquad \hat A_{ijk}=\hat A(r)\sqrt{\gamma}\epsilon_{ijk} \, .
\ee
The computation of the Gauss-Bonnet term gives,
\be \label{RGB}
R_\text{GB}= - 24 \left( \frac{1}{\rho^2} - \frac{\rho'^2}{\rho^2}   \right)  \frac{\rho''}{\rho} \, ,
\ee
while the Ricci scalar is still,
\be \label{Ricci}
R= 6 \left( \frac{1}{\rho^2} - \frac{\rho'^2}{\rho^2} - \frac{\rho''}{\rho}   \right) \, .
\ee
We can now write down the equations of motion. We obtain constant $\Lambda$ and $\theta$ on-shell, while the remaining equations can be written,
\begin{eqnarray}
3 M_\text{Pl}^2 \left( \frac{\rho'^2}{\rho^2}-\frac{1}{\rho^2} \right) &=& -(\Lambda+V(r)) \, , \label{deltaN}\\
M_\text{Pl}^2  \left( \frac{\rho'^2}{\rho^2}-\frac{1}{\rho^2}+2\frac{\rho''}{\rho} \right) &=& -(\Lambda+V(r)+\sigma_\text{w} \delta (r)) \, ,\qquad  \label{deltarho} \\
\frac{\sigma'}{\mu^4} A'(r) &=& \rho^3 \, , \label{deltaLambda}\\
\hat \sigma' \hat A'(r) &=& 24 \left(1-\rho'^2 \right) \rho'' \, . \qquad \label{deltatheta}
\end{eqnarray}
It should be noted that (\ref{deltaN}) and (\ref{deltarho}) are unchanged from General Relativity (GR), while (\ref{deltaLambda}) is the same as in \cite{KPS}. The potential,  
$$V(r)=\begin{cases} V_+ & r>0 \\V_- & r<0 \end{cases}$$
  is taken to be  a step function interpolating between the constant minima, whereas the bubble wall is modelled with a delta-function weighted by a tension $\sigma_\text{w}$.  
  
 Solving away from the bubble wall, we find that,
\be \label{rho}
\rho(r) = \frac{1}{q} \sin q(r_{0} + \epsilon r)\,,
\ee
where $\epsilon = \pm 1$, and,
\be \label{q2}
q^{2} = \frac{\Lambda +V}{3M_\text{Pl}^2}
\ee
 represents the local value of the spacetime vacuum curvature.  Here $q^2$ can be positive, zero, or negative for a spherical, planar or hyperbolic geometry respectively\footnote{Later, when we Wick rotate back to Euclidean signature, these will correspond to locally de Sitter, Minkowski and anti-de Sitter spacetimes.}. For the planar geometry, we can formally take the limit of \eqref{rho} as $q\to 0$, while for the hyperbolic case we analytically continue the formula to imaginary values of $q$. In all case, we can rewrite (\ref{RGB}) and (\ref{Ricci}) in terms of the local curvature $q$,
\be \label{q_squared}
R=12q^2 \, , \qquad
R_\text{GB}=24q^4 \, .
\ee
Matching conditions across the wall require continuity in $3$-sphere radius, $\rho$, and the $3$-form, $A$, at $r=0$, or in other words,
\be
\left[ \frac{1}{q} \sin qr_{0} \right]_{+}= \left[\frac{1}{q} \sin qr_{0} \right]_{-}, \qquad A(0^+)=A(0^-)
\ee
where labels $\pm$ denote evaluation in $\mathcal{M}_\pm$. In contrast, integrating  equations \eqref{deltarho} and \eqref{deltatheta} across the bubble wall yields the following discontinuities,
\be \label{discs_rho'}
2M_\text{Pl}^2  \frac{\Delta \rho'(0)}{\rho_0}=-\sigma_\text{w} \, , \qquad \Delta \hat A(0)= \frac{24}{\hat \sigma'} \left( \Delta \rho'(0) - \frac{\Delta \left (\rho'(0)^3 \right)}{3}\right)
\ee
where  $\Delta Q = Q_+-Q_-$ and $\rho_0=\rho(0^+)=\rho(0^-)$.  The jump in $\rho'$ is just the jump in extrinsic curvature  across the bubble wall, familiar from the Israel junction conditions \cite{Israel,Davis}.  Less familiar is the jump in $\hat A$, which can be rewritten as,
\be
 \hat A(0^+)-\hat A(0^-)= -\frac{12}{\hat \sigma'} \frac{\rho_0 \sigma_\text{w}}{M_\text{Pl}^2}\left[1-\left(\overline{\rho'(0)}\right)^2-\frac{\rho_0^2 \sigma_\text{w}^2 }{48M_\text{Pl}^4}         \right]
\ee
where $\bar Q=(Q_++Q_-)/2$ is the average across the wall. The jump in $\hat A$ occurs because $\hat A$ couples to energy-momentum through the curvature.  Tensional thin walls  therefore behave as membranes charged under $\hat A$, as in \cite{KPS}, although the mapping between the wall tension and the effective $3$-form charge is now different. In a physical set-up, we would, of course, expect the bubble wall to have finite thickness, allowing for a smooth but rapid transition in the value of $\hat A$.

Requiring that the bubble wall is supported by a sensible microscopic configuration, we require that it carries non-negative tension.  Through \eqref{discs_rho'} this places the usual kinematic constraint on the allowed configurations,
\be \label{tencon}
\Delta (\epsilon \cos qr_{0} ) \leq 0
\ee
Now let us turn our interest to the tunnelling rates between vacua. In the semi-classical theory of vacuum decay, including gravity, these rates are given by  \cite{Coleman_1,Coleman_2,Coleman_DeLuccia},
\be
\frac{\Gamma}{V} \sim e^{-B/ \hbar} \, ,
\ee
where,
\be
B=\delta S_\text{E} \myeq S_\text{E}|_\text{bounce}-S_\text{E}|_\text{initial vac} \, .
\ee
is the difference in the Euclidean actions for the bounce and the initial vacuum. Splitting $B$ into parts originating from different terms in the action, we can write,
 \be \label{B}
 B=B_\text{GR}-\sigma \delta c -\hat \sigma \delta \hat c \, .
 \ee
where $B_\text{GR}=-2 M_\text{Pl}^2 \Omega_3 \Delta  \left[\frac{1}{ q^2} [\rho'^3]^{0}_{r_\text{min}}\right]+\sigma_\text{w} \Omega_3 \rho_0^3$,  represents the tunnelling exponent computed in GR for the same geometrical configuration and $\Omega_3$ is the volume of the unit $3$-sphere. The flux terms are of the form,
 \begin{eqnarray}
\delta c & \myeq&\int_\text{bounce} F_4-\int_\text{initial vac} F_4 \nonumber\\
&=&-\frac{\mu^4}{\sigma'}\Omega_3 \Delta \left[\int_{r_\text{min}}^0 dr \rho^3 \right] \, , \nonumber\\
&=&  
-\frac{\mu^4}{\sigma'}\Omega_3 \Delta \left[-\frac{1}{3 q^4} [\rho'(3-\rho'^2)]^0_{r_\text{min}} \right ]\, .
 \end{eqnarray}
 and,
 \begin{eqnarray}
 \delta \hat c & \myeq &\int_\text{bounce} \hat F_4-\int_\text{initial vac} \hat F_4 \nonumber  \\
 & =&\frac{24 \Omega_3}{\hat \sigma'} \Delta \left[ 
   \rho'(r_\text{min})- \frac{1}{3} \rho'^3(r_\text{min})  \right]
   \end{eqnarray}
 Note that $\delta \hat c$ does not depend on quantities on the brane thanks to an exact cancellation that occurs due to the junction condition on $\hat A$. It is also worth highlighting that $r_\text{min}$ is \textit{a priori} different for the false vacuum and the bounce solution. Indeed, for  the bounce, the radial coordinate $r \in [r_\text{min}^-, r_\text{max}^+]$, passing from the interior, with $r<0$, to the exterior, $r>0$.  The precise values of $r_\text{max}$ and $r_\text{min}$ depend on the sign of the curvature and the orientation of the bubble \cite{KPS}:
  \be
 r_\text{min}=\begin{cases} -r_0 \, , & \epsilon=+1 \, , \\
  r_0-\frac{\pi}{q} \, , & \epsilon =-1, ~q^2 >0 \, , \\
  -\infty \, , &  \epsilon =-1, ~q^2 \leq 0 \, , \end{cases} 
 , \qquad
 r_\text{max}=\begin{cases}
  \frac{\pi}{q}-r_0 \, , & \epsilon =+1, ~q^2>0  \, , \\
  \infty \, , &  \epsilon =+1, ~q^2 \leq 0 \, , \\
   r_0 \, , & \epsilon=-1 \, .\\
 \end{cases} 
 \ee
Similarly, for the initial vacuum, the radial coordinate spans a range $r \in [r_\text{min}^+, r_\text{max}^+]$, although there is no longer any notion of exterior versus interior.

The contribution from the Gauss-Bonnet term in \eqref{B} is notable by its absence. Because of its topological nature in four dimensions, the bulk Gauss-Bonnet contribution is a total derivative, and is projected into a pure boundary contribution, at $r_\text{max}$ and $r_\text{min}$. These are then cancelled by the generalised Gibbons-Hawking boundary terms \eqref{boundaryterm}. 
 
 In principle, the constraint on the wall tension \eqref{tencon} does not forbid configurations in which the unbounded part of a Minkowski or AdS space tunnels to a new vacuum. However, these cannot be considered bubble solutions and are inconsistent with a suitable boundary prescription. 
The complete list of allowed transitions  are summarised in table \ref{table:configs2}.
\begin{table*} 
\tiny
\centering 
\begin{tabular}{| c || c | c| c| c |}
\hline
&$\text{S}_{+}-\text{S}_{-}$&$\text{S}_{+}-\text{H}_{-} $ &$\text{H}_{+}-\text{S}_{-} $&$\text{H}_{+}-\text{H}_{-} $ \\ 
\hline\hline
$\epsilon_{\pm} = 1$ & $(qr_{0})_{+} \geq (qr_{0})_{-} $ & \cmark& \xmark & $|q|_{+} \leq |q|_{-} $  \\  \hline
$\epsilon_{\pm} = -1$ & $(qr_{0})_{+} \leq (qr_{0})_{-} $ & \xmark & \xmark & \xmark   \\ \hline
$\epsilon_{+} = 1, \epsilon_{-} = -1$ & $\overline{ qr_{0} } \in [\pi/2, \pi]$ & \xmark &  \xmark&  \xmark \\ \hline
$\epsilon_{+} = -1, \epsilon_{-} = 1$ & $\overline {qr_{0} }\in [0, \pi/2]$ &  \cmark & \xmark  & \xmark \\
\hline
\end{tabular}  
\caption{  \label{table:configs2} Summary of allowed configurations. Those marked with a ``\cmark" are allowed while those marked with a ``\xmark" are not.  Note that $S$ denotes the sphere, $H$ the hyperboloid. Planar limits can be extracted  by taking $q_\pm \to 0$.} 
\end{table*} 
Focussing now on the allowed configurations  we note that they all have \cite{KPS},
\be
\rho'(r_\text{min})=1, \qquad -1 \leq \rho'(0^+) \leq \rho'(0^-) \, .
\ee
and so,
\begin{eqnarray}
B_\text{GR} &=& 2\Omega_3 M_\text{Pl}^2 \rho_0^2 \Delta \left[ \frac{1}{1+\rho'(0)} \right]  \geq 0\\
-\sigma \delta c &=& \Omega_3\frac{\mu^4 \rho_0^4}{3} \frac{\sigma}{\sigma'}{\Delta} \left[ \frac{1}{1+\rho'(0)} +\left( \frac{1}{1+\rho'(0)} \right)^2 \right] \quad \\
-\hat \sigma \delta \hat c &=& 0
\end{eqnarray}
Bringing it all together, we find that the tunnelling rate is given by an exponent,
\be
B=2\Omega_3 M_\text{Pl}^2 \rho_0^2 \left( 1+\frac{\mu^4 \rho_0^2}{6 M_\text{Pl}^2 } \frac{\sigma}{\sigma'}   \right){\Delta} \left[ \frac{1}{1+\rho'(0)} \right] 
+ \Omega_3\frac{\mu^4 \rho_0^4}{3} \frac{\sigma}{\sigma'} {\Delta} \left[ \left( \frac{1}{1+\rho'(0)} \right)^2 \right] \, ,
\ee
This suggests that a sufficient condition to avoid infinitely enhanced tunnelling rates, and a catastrophic instability in the theory,  is $\frac{\sigma}{\sigma'}>0$.

We now consider two special cases as in \cite{Coleman_DeLuccia}: tunnelling from de Sitter into Minkowski and tunnelling from Minkowski into Anti de Sitter. For tunnelling from de Sitter  into Minkowksi ($q^2 \to 0$), we have that  $\rho'(0^-)=1$ and $\rho'(0^+) \in [-1,1]$, and a tunnelling exponent, %
\be
B=B_\text{GR} \left[ 1+\frac{\mu^4}{12 q^2 M_\text{Pl}^2} \frac{\sigma}{\sigma'}s(8-3s)\right] \, ,
\ee
where, as in   \cite{Coleman_DeLuccia, KPS},  $B_\text{GR}=\Omega_3 \frac{M_\text{Pl}^2}{q^2} s^2$  and,
\be
s=1-\rho'(0^+)=\frac{\sigma_\text{w}^2}{2M_\text{Pl}^4 q^2}\left( \frac{1}{1+\sigma_\text{w}^2/4 M_\text{Pl}^4 q^2}\right) \, .
\ee
Given the constraint  $\frac{\sigma}{\sigma'}>0$ and the fact that in this case we have  $ s \in [0,2]$,  we see that the corrections due to OS  always suppress  this  tunnelling  event relative to GR.

Now consider the tunnelling from Minkowski into anti de Sitter ($0 \to -|q|^2$). Now we have   $\rho'(0^+)=1$ and  $\rho'(0^-) \geq 1$, and a  tunnelling exponent,
\be
B=B_\text{GR} \left[ 1-\frac{\mu^4}{12 |q|^2 M_\text{Pl}^2} \frac{\sigma}{\sigma'}s(8-3s) \right] \, ,
\ee
where, now,  $B_\text{GR}=\Omega_3 \frac{M_\text{Pl}^2}{|q|^2} s^2$ and,
\be
s=1-\rho'(0^-)=-\frac{\sigma_\text{w}^2}{2M_\text{Pl}^4 |q|^2}\left( \frac{1}{1-\sigma_\text{w}^2/4 M_\text{Pl}^4 |q|^2}\right) \, .
\ee
Transitions for which  $|q|^2< \sigma_\text{w}^2/4M_\text{Pl}^4$ are forbidden by energetic considerations  \cite{Coleman_DeLuccia}.  In anti de Sitter the bubble cannot get big enough for the energy stored in the wall to balance the energy stored in the interior. Once again, given the constraint  $\frac{\sigma}{\sigma'}>0$  and the fact that in this case we have  $ s \leq 0$, we see that OS corrections always suppress this tunnelling event.   To sum up, for a consistent theory of OS satisfying the constraint  $\frac{\sigma}{\sigma'}>0$, the allowed  inhomogeneous tunnelling events coincide exactly with those in GR, but always occur  at a slower rate. 

Finally we consider the evolution of the bubble once it has materialised. To see what it does, we simply  Wick rotate the bounce solution back to Lorentzian signature.  The Lorentzian solutions in our case are geometrically identical to those described in considerable detail, including their global structure, in \cite{KPS}. It is far too lengthy  to repeat here and we refer the reader to \cite{KPS} for further details.  The only difference in the generalised case under consideration here is the mapping between the local curvature and the fluxes.

To find this relation, we note that the integrated versions of (\ref{deltaLambda}) and (\ref{deltatheta}) are written as, 
\begin{eqnarray} \label{fluxes}
c &=& \int F_4 = \frac{\mu^4}{\sigma'}\int d^4x \sqrt{-g}=\frac{\mu^4}{\sigma'}(\Omega_++\Omega_-)   \\
 \hat c &=& \int \hat F_4 = -\frac{1}{\hat\sigma'}\int d^4x \sqrt{-g} R_{\text{GB}} =-\frac{24}{\hat \sigma'} (q_+^4 \Omega_++q_-^4 \Omega_-)
\end{eqnarray}
where $\Omega_+$ is the spacetime volume corresponding to the initial vacuum and $\Omega_-$ to the new vacuum. In particular, $\Omega_+$ includes the entire spatial volume  at all times up until the nucleation of the bubble, and then the exterior spatial volume afterwards. $\Omega_-$ is simply the bubble interior. 

Taking ratios of the two fluxes, we obtain,
\be
\frac{\Lambda_\text{flux}^2}{9M_\text{Pl}^4}=  \frac{q_+^4}{1+\mathcal{I}^{-1}}  +  \frac{q_-^4 }{1+\mathcal{I}} \, ,
\ee
where $\mathcal{I}=\frac{\Omega_{+}}{\Omega_{-}}$ is ratio of the spacetime volumes occupied by each particular vacuum, and we recall that $\Lambda_\text{flux}=\sqrt{-\frac{3}{8} M_\text{Pl}^4  \mu^4 \frac{\hat \sigma'}{\sigma'} \frac{\hat c}{c}} $.  From equation 
\eqref{q2}, we also have that,
\be
\Delta q^2 \myeq q_+^2-q_-^2=\frac{\Delta V}{3 M_\text{Pl}^2}
\ee
It follows that,
\be
q_\pm^2=\frac{1}{6M_\text{Pl}^2}\left[ -\Delta V(\mathcal{R}\mp 1) +\alpha \sqrt{(\Delta V)^2( \mathcal{R}^2-1)+4\Lambda_\text{flux}^2 }\right]
\ee
 where $\mathcal{R}=\frac{\mathcal{I}-1}{\mathcal{I}+1}$. Owing to the quadratic nature of the global constraint, our solution comes in two families, parametrised by  $\alpha=\pm1$. 

Now, if the old vacuum dominates the spacetime volume, then $\mathcal{I} \gg1$ and so $\mathcal{R} \approx 1$. It then follows that the local curvature in this region, $q_+^2$, is largely insensitive to the jump in vacuum energy, being given entirely by $\Lambda_\text{flux}$.  In contrast, $q_-^2$ is highly sensitive to $\Delta V$. The reverse is true when the new vacuum dominates the spacetime volume. Then we have $\mathcal{I} \ll1$ and so $\mathcal{R} \approx -1$: $q_+^2$ becomes highly sensitive to $\Delta V$, while $q_-^2$ is given by $\Lambda_\text{flux}$.

The computation of the spacetime volumes, which ultimately control which regions sequester vacuum energy most efficiently,  is a highly non-trivial exercise. The volumes are formally divergent to the infinite past and the infinite future. However the divergence rates can be correlated using the covariant junction conditions. Full details are presented in the appendix of \cite{KPS}, and the results can be carried over to the present case. We do so, however, with an additional word of caution. These ratios were computing using a global time regulator. Other regulators exist and could yield potentially different results due to the so-called measure problem, familiar from eternal inflation \cite{meas}. The global time regulator was chosen in \cite{KPS} because global coordinates cover the entire spacetime. We have nothing more to say on this difficult question. Let us simply quote the stated ratios and explore their consequences for the case under consideration here.

For a transition from $X$ to $Y$, where $X, Y$ are dS (de Sitter), M (Minkowski) or AdS (anti de Sitter), we label the corresponding volume ratio as $\mathcal{I}_{X \to Y}$. From \cite{KPS}, we then have, 
\begin{eqnarray}
\I_{dS \to dS} &\sim & \frac{q_-}{q_+} \, , \\
\I_{dS \to M} &=& 0 \, , \\
\I_{dS \to AdS} &=& \infty \, , \\
\I_{M \to AdS} &=& \infty \, , \\
\I_{AdS \to AdS} &=& \infty \, ,
\end{eqnarray}
The consequences of these ratios turn out to be the same as in \cite{KPS}, so we summarize those results. For phenomenologically interesting de Sitter to de Sitter transitions, we can have transitions in either direction. Transitions that lower the curvature ($q_- \ll q_+$) are far more probable and for these we have $\I \ll 1$, ensuring insensitivity to $\Delta V$ in the low curvature new vacuum. For the suppressed transitions that raise the curvature ($q_-\gg q_+$), we have $\I \gg 1$, again ensuring insensitivity to $\Delta V$ in the low curvature vacuum, although this time it is the old vacuum.  More generally, the following behaviour prevails: for a given transition, insensitivity to $\Delta V$ is achieved in the vacuum with lowest absolute curvature. The one exception to this rule is transitions from large curvature de Sitter to small  curvature anti de Sitter vacua. 
 
 This generic behaviour is important. It suggests that vacua with low absolute curvature do not require fine-tuning to achieve their low curvature: the sequestering mechanism will always take care of the required cancellations. We now see how this is common to all sequestering models.

\section{Conclusions} \label{sec:conc}

In this paper, we have explored the cosmological framework of {\it Omnia Sequestra}, the generalised theory of vacuum energy sequestering with the capacity to enforce cancellation of all  radiative corrections to vacuum energy, including both matter and graviton loops \cite{KP4}.   

As in older models of sequestering, the cosmological behaviour relies on certain historic integrals, although their structure is different in subtle but important ways.  As usual, the historic integrals feed into the residual cosmological constant that we observe through the large scale curvature. In OS, we find that there are potentially dangerous divergences coming from the singular region of spacetime. These represent a potential UV instability that could render the observed cosmological constant power law dependent on the UV cut-off of the theory. Such a scenario would mean a violation of naturalness and the theory would do no better than General Relativity.  However, it turns out that this behaviour can be tamed in a sufficiently large and old Universe, and eliminated altogether in a Universe that continues for eternity. For a Planckian cut-off,   92 more efolds in expansion will be sufficient.  We  also find that the scale of  residual cosmological constant can be assumed to be bounded above by the scale of the critical density today. This relies on two things: that the Universe grows old enough to tame any cut-off dependence in the historic integrals, and that the flux contribution is not too large. 

We also studied the effect of phase transitions through these historic integrals. For homogeneous transitions, we once again encountered potential naturalness problems that mirrored the UV sensitivity problem described in the previous paragraph. More precisely, we find that the residual cosmological constant at late times can become sensitive to  jumps in vacuum energy from transitions at early times.  Again, these contributions can be tamed as long as the Universe gets sufficiently old and eliminated altogether in an eternal universe. In particular, in a crude historical model, the effect of the QCD phase transition at high redshift  would require the Universe to continue for at least 23 more efolds. Again, with this proviso, we found that the late time behaviour became insensitive to the scale of the phase transiton.

The role of the 3-form fluxes was also investigated. This is boundary data, assumed to be UV insensitive and  taking on values that should be set empirically within the effective field theory. Nevertheless, there are geometric consequences of certain choices. In particular, we showed that for a vanishing flux ratio, the spatial geometry is forced to be  that of a hyperboloid.

The formalism for OS was reviewed in some detail in section \ref{sec:review}, and built upon to include the effect of spacetime boundaries. Owing to the non-trivial global dynamics in sequestering models, this extension is non-trivial but was important to allow for a study of inhomogeneous transitions, through the nucleation of a spherical bubble  and the bounce computation originally developed for GR by Coleman and De Luccia \cite{Coleman_DeLuccia}. Indeed, via a calculation of the  bounce, we were able to show that the allowed transitions coincided with those from GR.  An important new ingredient, however, was the mapping from the source potential to the local curvature. The local curvature became insensitive to the scale of the transition in the region of spacetime that dominated the volume. As in \cite{KPS}, the consequence of this is that generically those vacua with low absolute curvature are the least sensitive to the scale of the transition. This may seem obvious, but it is not. One could have a scenario in which the low curvature is highly sensitive to the transition scale and  one has to fine-tune. Indeed, there is one particular scenario where precisely this happens, although it is not generic.

The meaning of tunnelling probabilities in sequestering models may seem unclear at first glance, since the local value of the cosmological constant seems to have knowledge of whether or not tunnelling will occur.  Indeed, for a spacetime without any bubbles of true vacuum, there is complete cancellation of vacuum energy, whereas if a bubble exists to the future the cancellation is inexact, depending on the ratio of spacetime volumes as explained above. However, there is no  tension with the probabilistic interpretation of quantum tunnelling. On the one hand, the tunnelling rate per unit volume per unit time is faithfully captured by the bounce, corresponding to a saddle point of the {\it Euclidean} action.  The various spacetime configurations that may occur with and without bubble nucleation are all stationary points of the {\it Lorentzian} action.  This is exactly as in General Relativity, the only difference being that the sequestering solutions are also required to satisfy an additional global constraint. Furthermore, as a local observer, we have no way of knowing if the residual cosmological constant we measure contains contributions from inexact cancellations due to future bubble nucleation, or some future fluctuation in the local energy-momentum and its resulting contribution to the spacetime average.

Although our analysis has been thorough, some specific questions remain. In particular, we noted that the quadratic nature of Gauss-Bonnet ultimately means that there are multiple roots for the residual cosmological constant.  This deserves further investigation: does it lead to problems with well-posedness and branching; is there a physical mechanism for selecting one branch over the other? We have also been unable to attach any extra  physical significance to the generalised boundary conditions \eqref{Aij} we proposed for a well defined variational principle. Establishing this may yield a deeper understanding of the model and how it can be embedded in a more complete theory.

The presiding message is that all sequestering models exhibit similar cosmological behaviour. The phenomenology is consistent with observation, without fine-tuning, and seems to favour Universes that grow old and big. To a large extent, sequestering is best interpreted as a mechanism for cancellation of vacuum energy, rather than a specific model. With this perspective the future focus should really be to better understand how and why it does what it does, at a much deeper level. This depth of understanding should help facilitate the search for the mechanism at a fundamental level, probably as an emergent low energy effect in a UV complete theory.  

\begin{acknowledgments}
A.P. would like to thank David Stefanyszyn and Florian Niedermann for useful discussions, and Nemanja  Kaloper for the initial collaborations in vacuum energy sequestering. B.C. is supported by a STFC studentship and A.P. by a Leverhulme Research Project Grant and a STFC Consolidated Grant. A.P. would  also like to thank dead feline and boxing commentator, Andy Clarke, for advice on latin constructions used in the title.
\end{acknowledgments}


\begin{thebibliography}{99}

\bibitem{zeldovich} 
Y.~B.~Zeldovich,
JETP Lett.\  {\bf 6}, 316 (1967)
[Pisma Zh.\ Eksp.\ Teor.\ Fiz.\  {\bf 6}, 883 (1967)];
Sov.\ Phys.\ Usp.\  {\bf 11}, 381 (1968).

  
\bibitem{wein}
S.~Weinberg,
Rev.\ Mod.\ Phys.\  {\bf 61}, 1 (1989).

\bibitem{pol} 
  J.~Polchinski,
  hep-th/0603249.

\bibitem{cliff} 
  C.~P.~Burgess,
  arXiv:1309.4133 [hep-th].

\bibitem{me}
  A.~Padilla,
  arXiv:1502.05296 [hep-th].
  
\bibitem{KP1}
  N.~Kaloper and A.~Padilla,
  Phys.\ Rev.\ Lett.\  {\bf 112} (2014) 9,  091304.
  
  
\bibitem{KP2}
  N.~Kaloper and A.~Padilla,
  Phys.\ Rev.\ D {\bf 90} (2014) 8,  084023
   [Addendum-ibid.\ D {\bf 90} (2014) 10,  109901].
  
\bibitem{KP3}
 N.~Kaloper and A.~Padilla,
  Phys.\ Rev.\ Lett.\  {\bf 114} (2015) 10,  101302
  
\bibitem{KPSZ} 
  N.~Kaloper, A.~Padilla, D.~Stefanyszyn and G.~Zahariade,
  Phys.\ Rev.\ Lett.\  {\bf 116} (2016) 5,  051302
  
\bibitem{KPS}
  N.~Kaloper, A.~Padilla and D.~Stefanyszyn,
  Phys.\ Rev.\ D {\bf 94} (2016) no.2,  025022
  [arXiv:1604.04000 [hep-th]].
  
\bibitem{KP4}
  N.~Kaloper and A.~Padilla,
  Phys.\ Rev.\ Lett.\  {\bf 118} (2017) no.6,  061303
  [arXiv:1606.04958 [hep-th]].
  
\bibitem{etude}
  G.~D'Amico, N.~Kaloper, A.~Padilla, D.~Stefanyszyn, A.~Westphal and G.~Zahariade,
  JHEP {\bf 1709} (2017) 074
  [arXiv:1705.08950 [hep-th]].
 
\bibitem{irat}
  N.~Kaloper,
  arXiv:1806.03308 [hep-th].
  
\bibitem{mon}
  A.~Padilla,
  arXiv:1806.04740 [hep-th].
  
\bibitem{decap}
  A.~Adams, J.~McGreevy and E.~Silverstein,
  hep-th/0209226.
  
\bibitem{selftun}
  F.~Niedermann and A.~Padilla,
  Phys.\ Rev.\ Lett.\  {\bf 119} (2017) no.25,  251306
  [arXiv:1706.04778 [hep-th]].
  
\bibitem{lucas}
  L.~Lombriser,
  arXiv:1805.05918 [astro-ph.CO].
  
\bibitem{Demers}
  J.~G.~Demers, R.~Lafrance and R.~C.~Myers,
  Phys.\ Rev.\ D {\bf 52} (1995) 2245
  [gr-qc/9503003].
  




\bibitem{Lindeinf}
  A.~D.~Linde,
  Lect.\ Notes Phys.\  {\bf 738} (2008) 1
  [arXiv:0705.0164 [hep-th]].
\bibitem{Coleman_1}
  S.~R.~Coleman,
  Phys.\ Rev.\ D {\bf 15} (1977) 2929
   Erratum: [Phys.\ Rev.\ D {\bf 16} (1977) 1248].

\bibitem{Coleman_2}
  C.~G.~Callan, Jr. and S.~R.~Coleman,
  Phys.\ Rev.\ D {\bf 16} (1977) 1762.

\bibitem{Coleman_DeLuccia}
  S.~R.~Coleman and F.~De Luccia,
  Phys.\ Rev.\ D {\bf 21} (1980) 3305.

  


\bibitem{Boundary_terms}
  A.~Padilla and V.~Sivanesan,
  JHEP {\bf 1208} (2012) 122
  [arXiv:1206.1258 [gr-qc]].

\bibitem{Davis}
  S.~C.~Davis,
  Phys.\ Rev.\ D {\bf 67} (2003) 024030
  [hep-th/0208205].




\bibitem{GibbonsHawking}
  G.~W.~Gibbons and S.~W.~Hawking,
  Phys.\ Rev.\ D {\bf 15} (1977) 2752.
  doi:10.1103/PhysRevD.15.2752

\bibitem{O4_1}
  S.~R.~Coleman, V.~Glaser and A.~Martin,
  Commun.\ Math.\ Phys.\  {\bf 58} (1978) 211.
  
\bibitem{O4_2}
  A.~Masoumi and E.~J.~Weinberg,
  Phys.\ Rev.\ D {\bf 86} (2012) 104029
  [arXiv:1207.3717 [hep-th]].
  
\bibitem{Israel}
  W.~Israel,
  Nuovo Cim.\ B {\bf 44S10} (1966) 1
   [Nuovo Cim.\ B {\bf 44} (1966) 1]
   Erratum: [Nuovo Cim.\ B {\bf 48} (1967) 463].
  
   
  
  
\bibitem{GB}
  C.~Charmousis and A.~Padilla,
  JHEP {\bf 0812} (2008) 038
  [arXiv:0807.2864 [hep-th]].
  
  
  
\bibitem{meas}
  A.~Linde and M.~Noorbala,
  JCAP {\bf 1009} (2010) 008
  [arXiv:1006.2170 [hep-th]].

  
  
  

\end{thebibliography}
\end{document}